\documentstyle[prl,aps,twocolumn]{revtex}
\begin{document}
\draft
\title{Collective Charge Density Wave motion through an ensemble of Aharonov-Bohm rings}
\author{M. I. Visscher and B. Rejaei}
\address{Theoretical Physics Group, Department of Applied Physics/DIMES \\
Delft University of Technology, Lorentzweg 1, 2628 CJ Delft, The Netherlands}
\date{\today}
\maketitle

\begin{abstract}
We investigate theoretically the collective charge density wave motion through an ensemble of small 
disordered Aharonov-Bohm rings. It is shown that the magnetic flux modulates the threshold field 
and the magnetoresistance with a half
flux quantum periodicity $\Phi_{0}/2=h/2e$, resulting from ensemble averaging 
over random scattering phases of multiple rings. 
The magnitude of the magnetoresistance oscillations decreases 
rapidly with increasing bias. This is consistent with recent experiments on $\rm{NbSe}_3$ in 
presence of columnar defects [Phys. Rev. Lett. {\bf 78}, 919 (1997)].
\end{abstract}

One of the first experimentally observed quantum phenomena in charge density
wave (CDW) compounds is the oscillating magnetoresistance in ${\rm NbSe}_3$ in 
presence of small columnar defects, recently reported by Latyshev et al. 
\onlinecite{latyshev}. The collective response of the CDW to the
Aharonov-Bohm flux trapped inside the columnar defects reflected a $\Phi_{0}/2=h/2e$ periodicity. 
The occurrence of this period was related
to instantons\cite{bogachek} in ring-shaped 
commensurate CDW conductors, where large
scale quantum fluctuations of the CDW phase allow for macroscopic quantum
tunneling between degenerate ground states. 
Related work \cite{visscher} on CDW's in a ring geometry, however, predicts
a periodicity of $\Phi_{0}$ due to the modulation of the amplitude of the order parameter. 
It is doubtful whether these models apply to an array of columnar defects in a
planar film. Here we propose a different theoretical model, which is 
closer related to the actual experimental geometry, and can account for the
observed effects.

Figure \ref{fig:figa} shows schematically a planar film of one-dimensional CDW chains,
containing a small hole threaded by magnetic flux. The CDW is characterized by a 
complex order parameter $|\Delta|\exp(i\chi)$, where $|\Delta|$ is proportional to the
amplitude of the density modulation and the phase $\chi$ denotes its position. The size of the hole
is smaller than both the longitudinal $\xi_{\parallel}$ and transversal $\xi_{\perp}$ CDW coherence
lengths. As recently shown by Artemenko and Gleisberg \cite{artemenko1},
nonlinear screening of the phase distortions induced by a defect, leads to the 
formation of metallic islands surrounding the latter. Hence, in our model we
include a small conducting region around the hole where the CDW order is destroyed. In
the normal region, electrons can encircle the hole via
different paths, as indicated by the dotted lines, each picking up a
random scattering phase. Because of the strong anisotropy, the largest contribution
will arise from the middle chains. Therefore, as a model calculation, we proceed with a
one-dimensional treatment of the problem. 

Using microscopic equations 
for the sliding CDW motion ``over'' a general scattering
source, we show that the Aharonov-Bohm flux modulates the CDW threshold field. 
The periodicity $\Phi _0/2$ appears by ensemble averaging
over random scattering phases, as is known from the Al'tshuler, Aronov, and Spivak 
(AAS)-theory \cite{altshuler} for an ensemble of Aharonov-Bohm rings. Our results 
qualitatively account for the amplitude of the
magnetoresistance oscillations and its disappearance at higher bias.

In the framework of the kinetic equations \cite{artemenko}, the motion of
the quasi-particles and the condensate in a quasi one-dimensional CDW conductor can be
described by the semiclassical Green functions $g_{\alpha \beta
}^i(x;t,t^{\prime })$ where $i=\left\{ R,A,K\right\} $ and $\alpha ,\beta
=\left\{ 1,2\right\} .$ The retarded ${\bf g}^R$ and advanced ${\bf g}^A$
functions determine the excitation spectrum and the Keldysh component ${\bf g%
}^K$ describes the kinetics of the system. The subscripts 1 and 2 refer to
the right-, respectively, left-moving electrons at the two branches of the
linearized energy spectrum. We incorporate the effects of the Aharonov-Bohm
ring in this formalism, by imposing boundary conditions on the Green
functions.

The Green functions at the
right (R) and left (L) of a small scattering source of characteristic
length $l<\xi _0$ are related by
\begin{equation}
{\bf g}_R^i=({\bf M}^{\dagger})^{-1}{\bf g}_L^i{\bf M}^{\dagger},
\end{equation}
where ${\bf M}$ is the transfer matrix of the scatterer which satisfies the condition
${\bf M}^{-1}=\bbox{\sigma}_{3}{\bf M}\bbox{\sigma}_{3}$ in order to ensure
current conservation. The transfer matrix ${\bf M}$ can generally be paramtrized as
\begin{equation}
\label{tottrans}
{\bf M}=\left( 
\begin{array}{cc}
e^{i\eta _0}\sqrt{1+\lambda} & e^{i\varphi_0}
\sqrt{\lambda} \\ e^{-i\varphi_0}\sqrt{\lambda} & 
e^{-i\eta_0}\sqrt{1+\lambda}
\end{array}
\right),  
\end{equation}
where $\eta_{0}, \varphi_{0}$ are scattering phases and $\lambda/(1+\lambda)=R$ is the reflection probability of the scatterer 
\cite{footnote}. 
To investigate the motion of the CDW in the vicinity of the scatterer we proceed in five steps. First
we gauge away the phase $\chi
(x,t)$ of the CDW by performing the unitary transformation 
\begin{equation}
{\bf \tilde g}^i(x;t,t')={\bf U}^{\dagger }(x,t){\bf g}^i(x;t,t'){\bf U}%
(x,t^{\prime }),
\end{equation}
where ${\bf U}=\exp \frac i2\chi {\bbox \sigma }_3$. Then we apply a Fourier transformation with respect to the
time difference in Wigner's representation 
\begin{equation}
{\bf g}^{i}(x;t-t',T)=\int \frac{d\epsilon}{2\pi} {\bf g}^{i}(x,\epsilon,T)e^{-i\epsilon(t-t')/\hbar}, 
\end{equation}
with $T=(t+t')/2$. The third step is to restrict ourselves to the Keldysh Green function and 
integrate it over all energies. Using the identity $\int {\partial _\epsilon {\bf g}%
^Kd\epsilon }=4{\bbox \sigma }_3$ we obtain the boundary condition 
\begin{equation}
\label{boundy}
\bbox{\cal {L}}\int d\epsilon {\bf {\tilde g}}_R^K-\int d\epsilon {\bf {%
\tilde g}}_L^K\bbox{\cal {L}}+2\hbar(\dot \chi _R-\dot \chi _L)\bbox{\cal {L}}=0,
\end{equation}
where $\bbox{\cal {L}}={\bf U}_L^{\dagger }{\bf M}^{\dagger }{\bf U}_R$, and $%
\dot \chi _{R,L}=\partial _t\chi _{R,L}$ define the sliding CDW velocities at the right 
and left side of the scatterer, respectively. 
Next it is convenient to decompose
the matrices into the unit matrix and the three Pauli matrices as $\tilde{\bf g}%
=g_0{\bf 1}+{\vec g}\cdot \vec {\bbox{\sigma}}$ and $\bbox{\cal{L}}={\cal{L}}_{0}\bf{1}+\vec{\cal{L}}
\cdot \vec{\bbox{\sigma}}$, where $\vec g=(g_1,g_2,g_3)$, $\vec{\cal{L}}=({\cal{L}}_1,{\cal{L}}_2,{\cal{L}}_3)$
and $\vec {\bbox{\sigma}}=(\bbox{\sigma}_1,\bbox{\sigma}_2,\bbox{\sigma}_3)$%
. After substitution into (\ref{boundy}) we are left with the equations
\begin{mathletters}
\begin{equation}
\int d\epsilon(g_{0,R}^{K}-g_{0,L}^{K})+2\hbar(\dot \chi _R-\dot \chi _L)=0
\end{equation}
\begin{equation}
\label{vector}
{\cal{L}}_0 \int d\epsilon ({\vec{g}}_R^K-{\vec{g}}_L^K)+i
{\vec{\cal {L}}}\times \int d\epsilon ({\vec{g}}_R^K+{\vec{g}}_L^K)=0.
\end{equation}
\end{mathletters}
From the defenition of current
\begin{equation}
-I=\frac{eN(0)v_{F}}{4} \int d\epsilon {\rm Tr}{\tilde{\bf{g}}}^{K}+eN(0)v_{F} \hbar\dot{\chi},
\end{equation}
where $-e$ is the electron charge, $v_F$ is the Fermi velocity and $N(0)$ is the density of states at the Fermi energy,
it is seen that the first equation just expresses the conservation of current through the 
scatterer. In the quasi-stationary approximation, where we neglect the inertia of the
CDW $(\partial _t^2\chi =0)$ and the time dependence of the amplitude of the
order parameter, we know from the self-consistency equation that 
\begin{equation}
\label{selfcon}
\int d\epsilon g_1^K=0,\quad \int d\epsilon g_2^K=-i%
\frac{|\Delta|}{\gamma},
\end{equation}
where $\gamma$ is the dimensionless electron-phonon coupling
constant \cite{rejaei}. Substituting Eqs. (\ref{tottrans}) and (\ref{selfcon}) into Eq. (\ref{vector}),
we finally obtain
\begin{mathletters}
\begin{equation}
{\chi_{R}-\chi_{L}}=2\eta_{0},
\end{equation}
\begin{equation}
\label{motion}
\mu _R- \mu _L-\frac{|\Delta |}{
2\gamma}\sqrt{R}\cos(\chi-\varphi_{0})=0.
\end{equation}
\end{mathletters}
Here we have defined the electrochemical potential $4\mu=-\int d\epsilon g_{3}^{K}$, $\chi=(\chi_{R}+\chi_{L})/2$, 
and we have assumed equal amplitudes of the CDW order parameter at the
right and left contacts $|\Delta |_R=|\Delta |_L=|\Delta |$. The first equation represents 
the correlation between the CDW's on the right and left hand side
of the scatterer in equilibrium, and is necessary for a self-consistent treatment of pinning. The second equation is
the desired equation of motion for the CDW. It is similar to the
phenomenological single particle model for impurity pinning \cite{gruner}
with threshold potential $\mu_{T}$ given by $\mu_{T}=|\Delta|\sqrt{R}/2\gamma$. Equation (\ref{motion})
allows us to calculate the conductance of a CDW moving over a small-size but
arbitrary scattering source, which is characterized by its transfer matrix 
evaluated at the Fermi energy.

Now that we have studied the problem of a general scatterer we proceed by calculating 
the transfer matrix of a disordered ring
threaded by a Aharonov-Bohm magnetic flux. Following B\"uttiker \cite
{buttiker}, the two junctions of the ring are described by the unitary
scattering matrix ${\bf S}$, which relates the outgoing to the incoming scattering
amplitudes 
\begin{equation}
{\bf S}=-\frac 12\left( 
\begin{array}{ccc}
\Omega_{-}-\Omega_{+} & \sqrt{\beta} & \sqrt{\beta} \\ \sqrt{\beta} & 
-\Omega_{-} & \Omega_{+} \\ 
\sqrt{\beta} & \Omega_{+} & -\Omega_{-}
\end{array}
\right) ,
\end{equation}
where $\Omega_{\pm}=1\pm \sqrt{1-2\beta}$, and $0\leq \beta \leq \frac{1}{2}$ is a coupling
parameter describing the reflection at the entrance. 
We represent the upper $(+)$ and lower $(-)$ path around the ring with
the transfer matrices ${\bf N}_{+}$ and ${\bf N}_{-}$, parametrized as 
\begin{equation}
{\bf N}_{\pm }=\left( 
\begin{array}{cc}
e^{i\eta _{\pm }} & 0 \\ 
0 & e^{-i\eta _{\pm }}
\end{array}
\right),  
\end{equation}
where $\eta _{\pm }$ and $\phi _{\pm }$ are the scattering phases of the 
separate branches. For simplicity we consider here the clean limit without barriers in both arms, 
and associate only different scattering phases to different paths. A more general analysis will
be presented elsewhere\cite{visscher3}. The phase shift of the electrons due to the
magnetic flux is accounted for by the transformation ${\bf N}_{\pm
}\rightarrow \exp (\pm i\vartheta ){\bf N}_{\pm }$, where $\vartheta =\pi {\Phi }/{%
\Phi _0}$ with flux quantum $\Phi _0=h/e$.  After some algebra we obtain the
total transfer matrix {\bf M} of the disordered ring 
\begin{eqnarray}
\label{matrix}
 \Lambda M_{11} &=& \Omega_{-}^{2} \cos 2\phi + \Omega_{+}^{2}\cos 2\vartheta\nonumber \\
 &-& 4(1-\beta)\cos 2\eta
-4i\beta\sin 2\eta \nonumber \\
\Lambda M_{22} &=& -(\Lambda M_{11})^{*} \nonumber \\
\Lambda M_{12} &=&-\Lambda M_{21}=\Omega_{-}^{2}\cos 2\phi \nonumber \\
 &-& \Omega_{+}^{2}\cos 2\vartheta+4\sqrt{1-2\beta}\cos 2\eta,
\end{eqnarray}
where we have defined $\eta =(\eta _{+}+\eta _{-})/2$, $\phi =(\eta
_{+}-\eta _{-})/2$ and $\Lambda =4\beta\{ \exp (i\eta )\cos (\vartheta+\phi )-\exp
(-i\eta )\cos (\vartheta -\phi )\}$. It is easily verified that this transfer
matrix ensures current conservation. From Eq. (\ref{vector}) we obtain $\varphi_{0}=\pi/2$ 
and the self-consistency relation becomes
\begin{equation}
\label{phasedif}
\chi _R-\chi _L=2\arctan {( \frac{1-\beta}{\beta}\tan \eta -
\frac{\Omega_{+}^{2}\sin^2\vartheta+\Omega_{-}^{2}\sin^2\phi}{2\beta\sin 2\eta})} ,
\end{equation}
which oscillates as a function of the flux and the scattering phases $\eta$ and $\phi$. 
The threshold field also depends strongly on the magnetic flux and the scattering
phases through
\begin{equation}
\nonumber
\mu_{T}(\Phi,\eta,\phi)=\left| \frac{M_{21}}{M_{22}} \right| \frac{|\Delta|}{2\gamma}.
\end{equation}
This expression is dominantly periodic in the flux quantum, but also contains 
higher harmonics from weak-localization paths. As
expected for equal phases $\phi =0$, destructive interference at values $%
\Phi =\Phi _0/2\;{\rm mod}\;\Phi _0$ enhances the total backscattering and
thus the pinning force. For $\phi \ne 0$ its behavior is more complex.

So far we considered only a single Aharonov-Bohm ring. We now turn
to the problem of an ensemble of rings in a CDW system.
It is well known that the resistance of an ensemble of Aharonov-Bohm rings in series
or parallel retains only the half flux quantum periodicity $\Phi_{0}/2$ \cite{altshuler,murat}.
This is due to the different scattering phases of
subsequent rings. Interference effects of electrons traveling only half the
ring circumference average out. However, if electrons encircle 
the rings just once, the phase difference of time-reversed paths is $\Phi _0/2$ 
independent of the scattering properties of the
individual rings. In our model the ensemble average comes about from averaging 
over multiple columnar defects.  
In the limit of strong pinning where $\sqrt{R}|\Delta|/(\pi\hbar v_{F}n)\gg 1$, with large impurity potentials or 
a low impurity concentration $n$, it is known from the Fukuyama-Lee-Rice (FLR) model \cite{fukuyama} that the CDW adjust its 
phase to each defect in order to minimize the electrostatic Coulomb energy.   
As a consequence the net threshold field is proportional to the sum over 
all impurities. For uncorrelated defects in series the above formalism 
reproduces the FLR-model in this limit. If we assume a random distribution of scattering phases 
for the Aharanov-Bohm rings, we
obtain the net threshold field $E_{T}$ by taking the ensemble average over the scattering
phases $\eta $ and $\phi $ of a single hole 
\begin{equation}
\label{field}
E_{T}(\Phi)=\frac{1}{4\pi^2 ed} \int_{-\pi }^\pi d\eta \int_{-\pi }^\pi
d\phi \,\mu_{T}(\Phi ,\eta ,\phi ),
\end{equation}
where $d$ is the average distance between the columnar defects.
The result is shown in Fig. \ref{fig:figc} for different values of the parameter $\beta$. 
The average threshold field oscillates
as a function of half the flux quantum $\Phi _0/2$ around a constant value. With increasing 
reflection $\beta$ at the
junctions the total threshold shifts upwards and higher harmonics become visible 
due to the increasing dwell-time of electrons in the rings.
The appearance of a finite pinning force at zero field arising from the
columnar defects is consistent with experiments. 

Above we have shown that the threshold field of an ensemble of columnar defects oscillates 
with a half flux quantum period. 
To qualitatively understand the effect of the flux dependent threshold field
on the current-voltage characteristics, we distinguish between the high and low sliding velocity regimes.  
Near threshold the depinning and dynamics of CDW's can be described as a dynamical critical
phenomenon \cite{fisher}. The CDW current $I_{CDW}$ obeys the scaling relation
\begin{equation}
I_{CDW} \propto \left(\frac{E-E_T}{E_T}\right)^{\alpha},
\end{equation}
where the critical exponent $\alpha$ takes the value $3/2$ in mean-field \cite{myers}. 
Since the threshold field contains a flux-dependent term as in Eq. (\ref{field}), large oscillations in the 
magnetoresistance are expected with $\Phi_{0}/2$ periodicity. 
In the high velocity limit, however, the magnetoresistance oscillations disappear rapidly, since the CDW does
not see the pinning potentials and their flux dependence.
We remark that if we are allowed to neglect phase-slip processes, the potential drop
over a single defect should be smaller than the gap $|\Delta|$. The actual correlation between
the phases left and right is probably smaller that in our model calculation, due to, for example, the finite size of 
the hole and the metallic region. If we take into account the energy dependence of the
transfer matrix, it can be shown that the pinning force in Eq. (\ref{motion}) is then modified by 
 $|\Delta|\rightarrow |\Delta|\exp(-l/\xi_{0})$.

In comparison with the experiments \cite{latyshev}, this model accounts for the periodicity of the 
magnetoresistance oscillations and qualitatively for the decrease of the amplitude at large biases. The
observed minimum at zero field may arise from spin-orbit interaction, which is known
to determine the sign of the magnetoresistance oscillations in compounds with relatively large atomic-numbers 
\cite{altshuler}. 
This has not been taken into account presently and needs further investigation as well as an
extension to multiple channels. Experimentally, measurements of the dependence of the threshold field on the
magnetic flux should provide more
insights into the validity of our model.

We conclude by summarizing our results. We have derived microscopic equations for
the non-linear sliding CDW motion over an arbitrary scatterer. It is shown that magnetic flux trapped in a 
columnar defect modulates the threshold field. In a low density ensemble of defects, averaging results in
a half flux quantum periodicity of the magnetoresistance. The amplitude of oscillation decreases
rapidly with increasing bias.  

This work is part of the research program of the "Stichting voor
Fundamenteel Onderzoek der Materie (FOM)", which is financially supported by
the "Nederlandse Organisatie voor Wetenschappelijk Onderzoek (NWO)." It is a pleasure
to acknowledge useful discussions with Yuli Nazarov and Gerrit Bauer.

\begin{figure}
\caption{Schematic figure of a small columnar defect threaded by magnetic flux in a 
planar film of CDW chains. Electrons can encircle the hole through a metallic region, where
the CDW order is assumed to be destroyed (shaded area).}
\label{fig:figa}
\end{figure}

\begin{figure}
\caption{Averaged threshold field $E_{T}$ in units of $|\Delta|/2\gamma ed$ as 
a function of magnetic flux for different values of
$\beta=0.3; 0.35; 0.4; 0.45; 0.5$ from top to bottom.}
\label{fig:figc}
\end{figure}


\begin{references}
\bibitem{latyshev}  Yu. I. Latyshev, O. Laborde, P. Monceau and S.
Klaum\"unzer, Phys. Rev. Lett {\bf 78}, 919 (1997).

\bibitem{bogachek}  E.N. Bogachek, I.V. Krive, I.O. Kulik, A.S. Rozhavsky,
Phys. Rev. B {\bf 42}, 7614 (1990).

\bibitem{visscher}  M.I. Visscher, B. Rejaei and G.E.W. Bauer, Euro Phys.
Lett. {\bf 36} 613 (1996). 

\bibitem{artemenko1}  S.N. Artemenko and F. Gleisberg, Phys. Rev. Lett. {\bf %
75} 497 (1995).

\bibitem{altshuler} B.L. Al'tshuler, A.G. Aronov, and B.Z. Spivak, 
Pis'ma Zh. Eksp. Teor. Fiz {bf 33}, 101 (1981) [JETP Lett. {bf 33}, 94 (1981)];
B.L. Al'tshuler, A.G. Aronov, B.Z. Spivak, D.Yu. Sharvin, and 
Yu.V. Sharvin, Pis'ma Zh. Eksp. Teor. Fiz. {\bf 35}, 476 (1982) 
[JETP Lett. {\bf 34}, 588 (1982)].

\bibitem{artemenko}  S.N. Artemenko and A.F. Volkov, Zh. Eksp. Teor. Fiz. 
{\bf 80}, 2018 (1981) [Sov. Phys. JETP {\bf 53}, 1050 (1980)].

\bibitem{footnote} For the sake of simplicity we have taken $\bf M$ to be 
energy independent. $\bf M$ is then evaluated for particles at the Fermi surface.
This is a good approximation is the dwelling time of the particles in the
scatterer is small compared to $\hbar/|\Delta|$, which is the same as $l<\xi_{0}$.

\bibitem{rejaei} B. Rejaei and G.E.W. Bauer, Phys. Rev. B {\bf 54}, 8487 (1996).

\bibitem{gruner}  G. Gr\"uner, A. Zawadowski, and P.M. Chaikin, Phys. Rev.
Lett. {\bf 46}, 511 (1981).

\bibitem{buttiker}M. B\"uttiker,  Phys. Rev. B {\bf 32}, 1846 (1985).  

\bibitem{visscher3}  M.I. Visscher and B. Rejaei (unpublished).

\bibitem{murat} M. Murat, Y. Gefen, and Y. Imry, Phys. Rev. B {\bf 34}, 659 
(1986).

\bibitem{fukuyama} H. Fukuyama and P. Lee,  Phys. Rev. B {\bf 17}, 535 (1977). 
P.A. Lee and T.M. Rice, Phys. Rev. B {\bf 19}, 3970 (1979).

\bibitem{fisher} D.S. Fisher, Phys. Rev. Lett. {\bf 50}, 1486 (1983); Phys. Rev.
B {\bf 31}, 1396 (1985). 

\bibitem{myers} Other values for $\alpha$ are obtained in 
numerical studies, see C.R. Myers and J.P Sethna, Phys. Rev. B {\bf 47}, 11171 (1993).

\end{references}
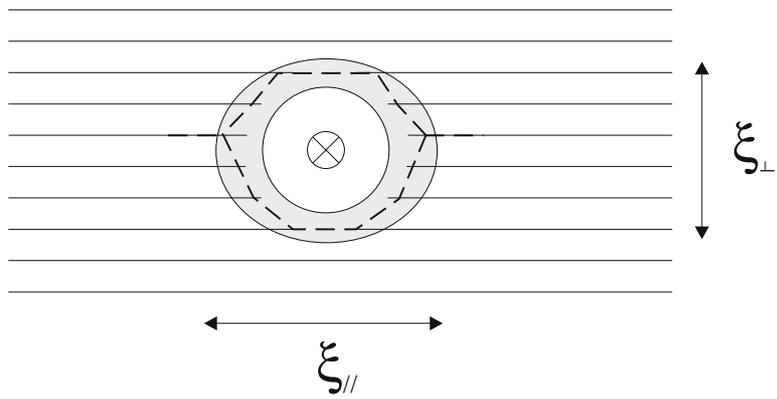
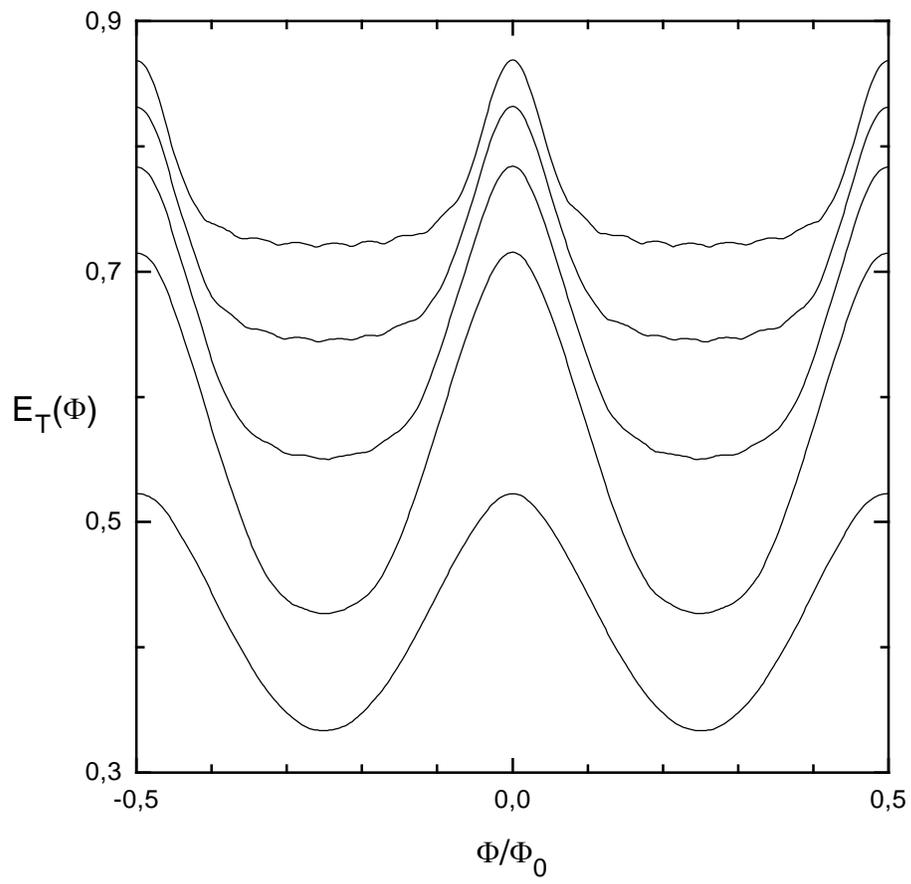
\end{document}